\documentclass[a4paper]{article}
\usepackage{fullpage}
\usepackage{amsmath}
\usepackage{amssymb}
\usepackage{graphicx}
\usepackage{relsize}
\usepackage{listings}
\begin{document}

\title{Casimir Forces via Worldline Numerics: Method Improvements
       and Potential Engineering Applications}
\author{Klaus Aehlig$^1$, Helge Dietert$^2$, Thomas Fischbacher$^1$, and Jochen Gerhard$^3$\\
  ${}^1$ University of Southampton, Faculty of Engineering and the Environment,\\
        Southampton, UK\\
  ${}^2$ University of Cambridge, UK\\
  ${}^3$ Frankfurt Institute for Advanced Studies,\\ Institut f\"ur Informatik, \\
Goethe Universit\"at Frankfurt, DE
}
\maketitle

\abstract{
\noindent
  The string theory inspired Worldline Numerics approach to
  Casimir force calculations has some favourable characteristics that
  might make it well suited for geometric optimization problems 
  as they arise e.g. in NEMS device engineering. We
  explain this aspect in detail, developing some refinements of the
  method along the way. Also, we comment on the problem of
  generalizing Worldline Numerics from scalars to photons in the
  presence of conductors.}

\section{Introduction}

Casimir forces, i.e., electromagnetic forces arising due to quantum
effects between uncharged conductors at distances much larger than
characteristic atomic radii, have first been predicted via theoretical
considerations in~1948~\cite{Casimir48}, and have since been verified
experimentally to 1\%~accuracy~\cite{PhysRevLett.81.4549}. While
initially a fringe subject -- albeit unfortunately one that drew
considerable attention from pseudoscientific authors -- research
interest in Casimir forces strongly increased in particular in the
last decade, and a number of novel theoretical tools were
developed~\cite{Scardicchio:2004fy,2003hep.th...11168M,Rahi:2010zr,
  PhysRevA.76.032106} that allow the study of the Casimir effect in
considerably more involved situations (geometries, material
properties, temperature) than earlier investigations. At present, one
major obstacle to research is that Casimir force calculations often
still are computationally very demanding. Nevertheless, the
development of theoretical tools and methods must go hand in hand with
progress in nanoscale manufacturing, for it is clear that a sound
understanding of the role of Casimir forces in nano machines will
become increasingly important as we learn to manufacture on shorter
length scales.

One approach to the calculation of Casimir forces is based on the
``Worldline approach'' developed by 
Gies, Klingm\"uller, Langfeld and Moyaerts~\cite{2002IJMPA..17..966G,2003hep.th...11168M,2006JPhA...39.6415G}.
While this mostly has been used to
study a simplified field theoretic model with massles scalars instead
of vector gauge bosons (photons), and nowadays alternative methods are
available to directly calculate electrodynamic effects even with
frequency-dependent optical properties of
materials~\cite{2009arXiv0904.0267R,2010PhRvA..81a2119M}, the
worldline approach is interesting for a number of reasons:

\begin{itemize}

\item Due to the probabilistic nature of the method, it is sometimes
  computationally comparatively cheap (depending on the geometry) to
  obtain a rough estimate of Casimir forces.

\item With very little effort, the calculation can be modified in such
  a way that it simultaneously gives all the forces on a number of
  bodies, making it potentially attractive for problems requiring a
  geometric shape optimization approach.

\item Finally, it is formulated in a way that is suggestive of a
  remarkably intuitive interpretation. This may manage to make this
  conceptually subtle quantum effect somewhat accessible to wider
  audiences not necessarily deeply familiar with quantum field
  theory. (One should compare this with the didactic problems related
  to the ``mesomeric effect'' in chemistry.)

\end{itemize}

\begin{figure}\label{fig:beam}
\begin{center}
\includegraphics[width=8cm]{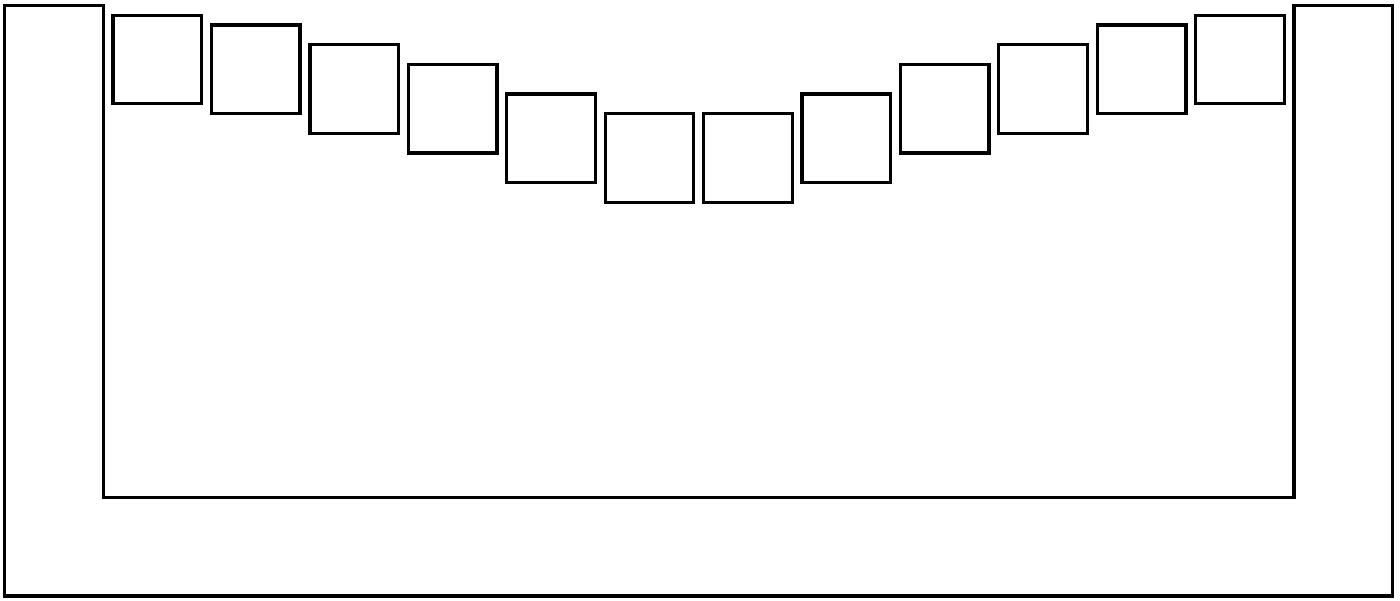}
\end{center}
\begin{caption}
{\scriptsize Beam bending under the influence of external forces and
  moments, a standard problem in Engineering, may be regarded as an
  energy minimization problem. Geometries such as the one shown above
  that involve bending beams at tens-of-nanometers length scales are
  presently being discussed as nanoelectromechanical memory
  devices~\cite{4160125}. Here, the beam also serves as a floating
  gate that allows electronic read-out of its mechanical state (upward
  bent or downward bent).  Discretizing the beam into elements with
  one translational degree of motion, the nature of the Worldline
  numerics algorithm makes it easy to \emph{simultaneously} calculate
  the Casimir forces on all elements with a computational effort less
  than one order of magnitude above the effort required to get a
  reliable estimate for the Casimir energy of the given configuration
  -- \emph{independent of the number of elements}. This is not easily
  achieved with a number of alternative methods to calculate Casimir
  forces.}
\end{caption}
\end{figure}

In this article, we will focus mostly on the second item in the list
above, which is elucidated in detail in
Section~\ref{sec:autodiff}. The key insight is that a deformable
object, such as a beam bending under the influence of Casimir forces
as shown in figure~\ref{fig:beam}, when discretized into $N$~elements
(blocks) that are partially restricted in their relative motion, may
require a certain computational effort, $T$~processor-seconds, for a
reasonable estimate of the Casimir energy of a given configuration,
but will then always allow the simultaneous computation of Casimir
forces (and moments) on \emph{all} $N$~elements in a way that requires
at most~$5T$ processor-seconds, \emph{irrespective of the number of
  bodies~$N$}.

\section{Computational Methods}

In contradistinction to other approaches, Worldline Numerics is
remarkably simple to implement in a computer program, requiring almost
no advanced software library infrastructure to deal with issues such
as parallel sparse matrix linear algebra, multipole moments, or large
matrix eigenvalues. Still, there are a number of useful improvements
of the basic method that help to ensure making effective use of
computational resources. As these are often linked to one another, it
makes sense to address them in a coherent fashion. 

\subsection{Monte Carlo integration over Loops}

The Casimir energy for a static geometry that can be modeled by
a position-dependent potential~$V(x)$ is given as the quantum
effective action per unit time:
\begin{equation}
E_{\rm Casimir}=\frac{\Gamma[V]}{\int_{\tau=\tau_-}^{\tau_+}\,d\tau}
\end{equation}

The ``Worldline Numerics'' approach by Gies, Langfeld, and
Moyaerts~\cite{Gies:2003} is based on re-writing the logarithm in the
effective action as an integral a la $\ln
(p/q) \rightarrow \int_0^{\infty} \frac{dT}{T}\left(\exp(-px)-\exp(-qx)\right)$
and re-expressing the operator trace as a Feynman path integral.  This
leads, for a real scalar field of mass~$m$ interacting only with the
external potential~$V\!$, to an expression for the effective action that
is numerically tractable via Monte Carlo methods. The key expression
from~\cite{Gies:2003}, which we repeat here for the convenience of the
readers, is:
\begin{equation}\label{eq:effective-action}
\Gamma_\Lambda[V] = - \frac 12 \frac 1{(4\pi)^{2}}\int_{1/\Lambda^2}^\infty
\frac{dT}{T^{3}} e^{-m^2T}\int d^4x
\left[\left\langle W_V[y;x,T]\right\rangle_y - 1\right]
\end{equation}
where an UV cut-off regulator~$\Lambda$ has been introduced. 
Here, the expectation value~$\langle\cdot\rangle_y$ is the ensemble
average over all closed loop (c.l.) gaussian random 
walks~$y:[0;1]\mapsto\mathbb R, y(0)=y(1)$ of 
``Wilson loops'' re-scaled to proper time~$T$. 
Let the statistical weight of the loop~$y$ be
\begin{equation}
p[y]=\exp\left(-\int_{t=0}^{t=1}dt\,\dot y(t)^2/4\right)
\end{equation}
then:
\begin{equation}
\left\langle W_V[y;x,T]\right\rangle_y=
\frac{\int_{y\;\rm c.l.}\mathcal{D}y\,W_V[y;x,T]\,p[y]}
{\int_{y\;\rm c.l.}\mathcal{D}y\,p[y]}
\end{equation}
where $W_V$ depends on the \emph{path} $y$ and on position 
shift~$x$ and propertime~$T$:
\begin{equation}
W_V[y;x,T]=\exp\left(-T\int_{t=0}^{t=1}dt\,V(x+\sqrt{T}y(t))\right).
\end{equation}

From this expression for the effective action of a free scalar
interacting with a potential, Casimir forces can be obtained by using
the position dependency of the potential~$V(x)$ to model the geometry,
and calculating energy changes associated with changes to the geometry.

While the applicability of this model for the calculation of real
Casimir forces is questionable (even for perfect conductors) as the
physics of photons is quite different from that of a scalar field, the
remarkable conceptual simplicity of the above expressions certainly
warrants a deeper investigation of its properties and potential
utility, for it might actually allow a (yet undiscovered)
generalization to the photon case. For the electromagnetic case, one
would naturally want to start with investigations of perfect conductor
surfaces, and the (obvious) scalar pendant of this is a
potential~$V(x)$ that suppresses all quantum
fluctuations~\emph{inside} the given bodies. It is not difficult to
see that one may alternatively restrict the potential to have non-zero
values only close to the surfaces of bodies, taking
\begin{equation}
  V(x) = \lambda \int_\Sigma d^2\sigma \delta^3(x-x_\sigma) 
\end{equation}
and considering the limit $\lambda\to\infty$. Then,~$W_V[y;x,T]$ reduces to:
\begin{equation}
 \exp\left[-T\int_0^1 dt V(x+\sqrt T y(t))\right]
  =
  \begin{cases}
    1 & \text{Loop pierces a surface}\\
    0 & \text{Loop does not pierce a surface.}
  \end{cases}
\end{equation}

Substituting $s=\sqrt{T}$ to eliminate the square root, and using
translation invariance of the integral to ensure all loops have center
of gravity at the origin, the expression for the geometry-dependent
regularized Casimir energy then is:
\begin{equation}
\Gamma_\Lambda = -\frac{1}{(4\pi)^2} \int_{1/\Lambda}^\infty \frac{ds}{s^5} \langle\Theta_V(sy,x)\rangle_y
\end{equation}
where the mean value is over all unit loops, $sy$ is the loop
$y(t)$ scaled pointwise around its center of gravity~$\bar y$ 
by the factor $s$, and
\begin{equation}
 \Theta_V(sy,x) =
  \begin{cases}
    0 & \text{Loop does not pierce a surface}\\
    1 & \text{Loop pierces a surface.}
  \end{cases}
\end{equation}

The problem with this approach is that the Casimir energy attributed
to the surface of any single body goes to infinity as we send the
energy regulator~$\Lambda$ to infinity, due to the contribution from
very short loops close to the surface. (This is, of course, a
non-physical artefact related to the ``geometry much larger than
atomic scales'' approximation.) In order to predict Casimir forces
between different objects we are only interested in the dependency of
the energy on the relative position of these objects. Therefore, it
makes sense to modify this scheme in a way that (i) the contribution
of each loop is taken into account relative to a configuration in
which all objects are at infinite separation from one another, and
(ii) the Casimir energy contribution attributed to loops piercing only
one object surface (hence, ``belonging'' to that object) is taken as
zero.

Consider $n$ objects with potentials $V_1,\dots, V_n$. Then the total
potential is $V=V_1+V_2+\dots +V_n$ and we use the freedom to shift
the absolute energy level to define the ``interaction Casimir energy''~$E$ 
as in~\cite{Gies:2003} as the energy difference relative to a configuration
in which every body is at an effectively infinite distance from every 
other body:
\begin{equation}\label{eq:subtraction}
  E = \left(\Gamma[V] - \Gamma[V_1] - \Gamma[V_2] - \dots - \Gamma[V_n]\right)/\Delta \tau
\end{equation}

Using this, we get
\begin{equation}
  E =  -\frac{1}{(4\pi)^2}\int_{1/\Lambda}^\infty \frac{ds}{s^5} \langle\Theta(sy,x)\rangle_y
\label{eq:loopenergy.final}
\end{equation}
with $\Theta$ given by
\begin{equation}
  \Theta(sy,x) =
  \begin{cases}
    0 & \text{Re-scaled loop does not pierce any surface} \\
    1-n & \text{Re-scaled loop pierces the surfaces of $n\ge 1$ objects}
  \end{cases}
\end{equation}

If~$n$ objects come close to one another, every loop that pierces all
of them can be regarded as the image of $n$~loops, each to be
considered as being attached to (and moving with) that body. Hence,
when objects are in proximity, we count a loop \emph{once} that would
have been counted $n$~times instead for separated objects. (Note that
the counting weight of both a loop that pierces no surface, and a loop
that pierces only one surface, is zero.) If the objects are now
spatially separated the integral $\int_{1/\Lambda}^\infty
\frac{ds}{s^5} \langle\Theta(sy,x)\rangle_y$ is finite and well
behaved for $\Lambda\to0$, so we can safely set $\Lambda=0$.  One is
easily convinced that this is indeed the correct expression by
considering a simple geometry (such as two parallel flat slabs) and
requesting that the Casimir force does not change if one object is
instead thought of as being made of two adjacent bodies. The counting
weights are dictated by the convention for the ``zero energy''
configuration.

For each loop~$y$, the weight~$\Theta$, as a function of the
re-scaling factor~$s$, is piece-wise constant. The~$s$-integral hence
can easily be performed analytically. Rather than being only a
convenient simplification that saves computing time, this property
plays a crucial role for the efficient simultaneous computation of
multibody forces, cf. Section~\ref{sec:autodiff}.

\subsection{Loop Generation}\label{subsubsec:generation}
When trying to evaluate Equation~\ref{eq:loopenergy.final}, one
naturally would try to discretize the loop as consisting of a finite
number of straight sections. Taking the procedure literally, the
presence of complicated curved geometries would mandate
computationally fairly expensive ray-surface intersection checks. In
many cases, a better investment of the computational effort may be to
instead make the number of discretization points on the loop
sufficiently large to ensure that simple inside/outside checks applied
to each point give a reasonably close approximation. Still, generic
ray/surface intersection checks can become useful, especially if the
complicated multiple integral in Equation~\ref{eq:loopenergy.final}
(over loop shapes, loop sizes, and loop centers of gravity) can
partially be evaluated by analytic, or rather semi-analytic means that
involve numerical approximation of integral boundaries. This is
relevant for the discussion in Section~\ref{sec:scaling}, and may make
major computational improvements of the method possible. In this
context, we want to point out the existence of advanced algorithms
useful for ray/surface intersection problems, such as Comba and
Stolfi's Affine Arithmetic~\cite{CombaStolfiSiggraph93,Knoll09}.
%
%

In order to generate a properly distributed random sample of loops, we
first generalize the problem to finding a process that produces
piecewise straight paths with gaussian length distribution (of given
standard deviation) for a given starting and end point (not
necessarily coincident). This problem can be re-phrased as finding
pairs of such paths, each with its own starting point but at first
without any constraint on their endpoints, and imposing the condition
that they meet at their endpoints. Concatenating the first path to the
reverse of the second solves the problem of finding a path with the
correct distribution between two given points. One easily sees that
the distribution of the midpoint is still gaussian (being the product
of two gaussian distributions). Hence, we can sample a loop by
recursively sampling an intermediate point in the interval between a
given start and end point.

This method, known as the~``d~(`doubling') loop
algorithm''~\cite{Gies:2005sb}, manages to generate closed loops with
the desired distribution with very little effort. A slight problem of
this approach is that it only generates loops for which the number of
vertices is a power of two, but as one usually is only interested in
getting below a given resolution, and this method will at worst
require an effort too large by a factor~2, this is practically
irrelevant. Nevertheless, the reasoning presented above is readily
generalized to non-equal subdivisions (e.g. 1/3 plus 2/3), and from
there to also allow recursive subdivision into other numbers of parts,
but this is typically not needed. In order to generate a loop
containing 10 points using the extended d-loop method, one would first
determine the opposite point (point \#6) of the starting point
(point \#1), then for both arcs use a $2:3$ split of the associated
variance budget and work out the gaussian distribution of the
corresponding intermediate point's position. Sub-division of the arc
of length~2 is straightforward, while the length~3 arc would first be
split~$1:2$ using the same approach.

Rather than choosing starting points randomly in space and then
determining the location of the halfway-round-the-loop point, it makes
sense to perform stratified spatial sampling on a lattice. To do so,
we choose the first point to be the lattice point and take the halfway
point to be gaussian distributed with mean the first point and
standard deviation a characteristic length. We then continue sampling
a loop of that length and later scale it around the midpoint between
the first point and the halfway point appropriately to obtain a unit
loop. In that way, the midpoint is gaussian distributed around the
lattice point with a given length scale.

If we want unbiased integration by taking loops for each lattice
point, the lattice points have to be representative
for loops sampled in their vicinity, with a characteristic length
being that of the grid. This is certainly true if the grid is fine
compared to any characteristic length of the geometry. However, the
same can be achieved for arbitrary grids, if we take the
characteristic length in the just described stratified sampling to be
that of the grid.

A different kind of lattice effects has to be taken into account for
methods computing a force as a difference in energy for two given
geometries. Such methods would typically put a fixed set of loops on
each lattice point and add up their energy contributions. Then they
would do the same for the same geometry with one object moved in a
particular direction. The difference in energy is then proportional to
the force component on the moved object in the given direction.

To focus on the net effect, as we do with symmetries
(see~\ref{subsubsec:cancellation}), one typically would use the same
set of loops for both geometries. Also, to have for each loop a
corresponding shifted loop, the amount the object is moved has to be a
multiple of the grid length (in this direction). While doing otherwise
would not necessarily yield a bias, doing so significantly improves
the convergence speed of this method.

For our purposes, we typically only ask simple questions about each
loop, such as ``which objects does it hit?'', or (at most) ``For what
scaling intervals does this loop, centered at $x_{cm}$ but re-scaled
in size, hit object~$O_n$?''. In order to answer these, only very
little information needs to be stored when visiting the loop point by
point. So it is possible to implement the relevant algorithms in such
a way that the loop is generated on the fly, and we never have to
store the entire loop in memory -- the number of points we have to
remember is about the binary logarithm on the loop length. This yields
an algorithm with a very small memory footprint and attractive
characteristics for computing architectures that emphasize a high
degree of parallelism between very simple cores.

\subsection{Numerical Integration over the Scaling Factor}
\label{subsec:directintegration}
One approach to obtain energies---and so ultimately, by comparing
energies for different geometric configurations, forces---is to
directly evaluate the integral in Equation~\ref{eq:loopenergy.final}
numerically. Naively, one would have to, for various values of~$s$,
estimate $\langle\Theta(sy,x)\rangle_y$ and summing up. Since the
order of summing up does not matter, we can as well compute the
expectation value of the following process.
\begin{quote}
Choose $s$ uniformly at random from the interval $[a,b]$. Randomly
generate a loop~$sy$ of size~$s$ and count~$(1-n)/s^5$ if the loop
hits~$n\geq 2$ objects, and~$0$ otherwise.
\end{quote}
Here $[a,b]$ is an interval big enough so that integrating over that
interval does not differ noticeably from integrating over all 
positive reals.

Looking at that random process more closely, one notes that the
information about the random loop we use is the number of objects it
hits. We have to pay particular attention to short loops that are just
long enough to barely touch multiple objects, for they give the
largest contribution to the sum. One should note that it is not
possible to attribute a useful physical meaning to absolute
differences in loop scaling factors~$s$: for a loop that hits (at
least) two objects, the effect of changing~$s$ to~$s+0.1$ very much
depends on what the magnitude of~$s$ is. As relative changes of the
scaling factor hence are more important than absolute changes, we much
prefer a distribution, when sampling loops, that handles all orders of
magnitude equally. In other words, we prefer a distribution where the
logarithm of~$s$ is uniformly distributed on $[\ln(a),\ln(b)]$.
 
When changing the distribution of~$s$, we also have to transform the
weight attributed to each sample accordingly.  Taking the logarithm
of~$s$ to be uniformly distributed, rather than~$s$ itself, each
value~$s$ will be~$1/s$ times as likely as before. To still get the
same expectation, we have to multiply each value by~$s$. Hence, we are
finally left with estimating the expectation of the following process:

\begin{quote}
Chose $\sigma=\ln s$ uniformly at random in the interval
$[\ln(a),\ln(b)]$. Randomly generate a loop of size $e^\sigma$ and count
$(1-n)e^{-4\sigma}=(1-n)s^{-4}$ if loop hits $n\geq 2$ elements, 
and $0$ otherwise.
\end{quote}

\subsection{Symbolic Integration over the Scaling Factor}
\label{sec:scaling}
It makes sense to try to perform at least part of the integration
needed to evaluate~eq.~\ref{eq:effective-action} symbolically, for two
independent reasons. While this may on the one hand help to simplify
the problem, it also gives us a much more useful handle on problems
that involve changing geometries. As we are much more interested in
Casimir forces (and moments) than just energies, this is obviously
desirable.

In particular, we can, as in~\cite{Gies:2006cq}, typically perform the
integration over the loop scaling factor
$\int_{s=0}^\infty \frac{ds}{s^5} \Theta(sy,x)$ symbolically.

If we have sampled a loop $y$, we can compute for each sampling point
the values of~$s$ for which this sampling point is inside a given
object. Often, this is just an interval, or at worst the union of a
few intervals. By merging these intervals for each sampling point, we
can compute the set of~$s$ values for which the loop hits the given
object. Now, as~$\Theta$ counts the number of objects hit by the loop,
it is piecewise constant on the partitioning so obtained; 
if~$\Theta=n$ for~$T\in[a,b]$, we 
have~$\int_a^b \frac{ds}{s^5}\, \Theta(sy,x) = n(a^{-4}-b^{-4})/4$.

Note that this means that we also do not need to specify the region
of~$s$ which we want to sample, i.e. our method does not need to know
a geometric length-scale.

\subsection{Forces on Multiple Bodies via Sensitivity Backpropagation}
\label{sec:autodiff}
In order to calculate forces, we have to determine by how much Casimir
energies change when changing the geometry. Taking the
limit~$\Lambda\to\infty$, the subtraction
scheme~eq.\ref{eq:subtraction} is only compatible with geometry
changes that change the position and orientation, but not shape, of
individual bodies: evidently, the Casimir energy attributed to a
single body via this scheme is zero, regardless of its shape. As we
naturally would expect the Casimir force between two parallel flat
plates to not disappear when connecting them with a thin wire (so that
they become a single object), the regularization prescription that
amounts to attributing forbidden loops to specific objects cannot be
compatible with (unconstrained) shape changes.

Even if we limit ourselves to shifts and rotations of bodies,
retaining~eq.~\ref{eq:subtraction}, it makes sense to describe geometry
changes in a more general way. We hence consider the potential~$V$ to
be a function of multiple geometry parameters (positions and angles), 
e.g.
$V=V(r^a_1,r^a_2,r^a_3,r^b_1,r^b_2,r^b_3,\alpha^a_1,\beta^a_2,\gamma^a_3;\ldots)$.

If we perform integration over the scaling parameter~$s$ analytically
for every loop and consider the intervals over which the subset of
objects pierced by that loop does not change, then the interval
endpoints, will become analytic functions of the geometry parameters
-- and so will the loop's contribution to the total Casimir energy.

For any given loop, we have to perform a fairly simple computation,
which in the end gives a single number. Furthermore, it is perfectly
feasible to design the algorithm in such a way that at the end of the
computation of the loop's contribution to the Casimir energy, we still
remember all the intermediate values that entered that calculation. In
such a situation, there is a standard method (in the sense of an
algorithmic transformation on the program that calculates the loop
energy) that allows a fast evaluation of the gradient with respect to
all the geometry parameters. Irrespective of the number~$N$ of such
parameters, and the complexity of the intermediate expressions, it is
possible to obtain the gradient to full numerical accuracy with at
most $5$~times the computational effort needed to calculate the scalar
function (often even much less). In comparison, the naive direct
method of calculating the gradient by comparing function values would
require at least~$N+1$ full evaluations of the function (and then only
give a result with reduced numerical accuracy).

The generic approach that makes this possible, which has become known
under the names ``automatic/algorithmic differentiation'',
``sensitivity backpropagation'', or ``adjoint
code''~\cite{Speelpenning80,Griewank89onautomatic} is essentially
based on this idea:

\begin{itemize}

\item Every intermediate result used in the calculation gets stored away              for later use (i.e. none may be dropped or over-written).

\item To each such intermediate quantity~$I_k$ stored, we associate a 
      buffer that can store another number~$\bar I_k$, initialized to
      zero at the beginning of the program. Ultimately, these will
      each end up holding the answer to the question: ``If, at the
      point when~$I_k$ became first known during the calculation of
      the function value, we interrupted the computation, changed that
      value from $I_k$ to $I_k'=I_k+\epsilon$, and then allowed the
      computation to proceed without further modifications now
      using~$I_k'$ instead of~$I_k$, by how much would the final
      result then change, relative to~$\epsilon$ (in the limit of
      small~$\epsilon$)?''. Hence, the~$\bar I_k$ will eventually
      become \emph{sensitivities} that describe the dependence of
      the result on the given intermediate quantity.

\item Treating input parameters in the same way as intermediate values,
      the sensitivities on all the input values give the function's
      gradient.

\item Sensitivities are calculated in a two-step process: first,
      the function is evaluated once to obtain all the intermediate
      quantities for the given choice of input parameters. Then,
      starting from the result, which has been obtained from an
      arithmetic operation involving intermediate values,
      sensitivities for the last intermediate get updated. As these
      again have been the result of some arithmetic operation, the
      sensitivities for the intermediate values they have been
      obtained from can be determined, etc. As one intermediate
      quantity may be used multiple times throughout the calculation,
      it is important to collect incremental contributions to 
      sensitivities when going backwards through the computation.
\end{itemize}

The theoretical maximum effort factor of~$5$ can be traced back to the
effort required to handle multiplication/division of intermediate
values. Evidently, if the sensitivity of the result on the
intermediate quantity~$I_k$ is known in~$\bar I_k$, and $I_k$ was
obtained as the sum of~$I_i+I_j$, then the sensitivities of~$I_i$ and
$I_i$ must be increased by~$\bar I_k$ (for if e.g.~$I_j$ gets used
multiple times, then~$\bar I_j$ receives multiple increments). If,
however,~$I_k$ is the \emph{product} of~$I_i$ and~$I_j$, then~$\bar
I_i$ must be increased by~$I_j\cdot \bar I_k$ and vice versa -- the
combined read/add/store operations give rise to a bounded
multiplicative factor for the total effort.

In practice, the corresponding calculations are even considerably
simpler than what the general theory of algorithmic differentiation
suggests, as the partial derivatives $\partial E/\partial g_j$~with
respect to the geometry parameters~$g_j$ are, at least for simple
surfaces, obtainable directly as a by-product of the forward
calculation. Comparing the Worldline method with the remarkable
approach discovered by Rodriguez, Ibannescu, Iannuzzi, Capasso,
Joannopoulos, and
Johnson~\cite{2007PhRvL..99h0401R,2007PhRvA..76c2106R}
or for that matter any method that involves numerically solving
discretized sparse linear operator equation systems, one would
naturally not expect sensitivity backpropagation to be as readily
applicable with these approaches as with Worldline Numerics. On the
one hand, sensitivity backpropagating an iterative linear solver would
require remembering the intermediate values from all iteration steps
(otherwise thrown away), and also, the calculation may have happened
to produce a solution before having explored some of the dependencies
sufficiently well. While research has been done on backpropagating
linear solvers, the incorporation of this strategy into Worldline
Numerics undeniably is much easier to accomplish.

\subsection{Adaptive Sampling}\label{subsec:adaptive}
In a typical geometry, essentially the whole energy or force is
contributed by few, comparably small regions. These are typically the
regions where two objects come closely together.

While we still have to sample loops in such a way that we integrate
over all of the relevant region of space, it is worthwhile to focus
effort mainly on these highly contributing areas, as the absolute
uncertainty of our Monte-Carlo estimation is much higher there. We
achieve this in the following way: We first specify an absolute
accuracy to which we want the density estimated to at every
point. When later sampling the density at a given point, we first take
a specified minimum of samples. From that we estimate the (unbiased)
variance of our sampling at this point. We continue sampling until a
pre-defined (95\%) confidence interval for the sampling mean is
smaller than the prespecified accuracy.

\subsection{Living with Cancellation}\label{subsubsec:cancellation}
Some geometries, like the ``cylinders with sidewalls'' geometry
studied in~\cite{2008PhRvA..77c0101R}, show a high
degree of symmetry. While perfect symmetry helps to reduce the
computational effort as the calculation can be restricted to a
fundamental domain, slightly non-symmetric configurations often are a
problem if we want to compute the force on an object that gets pulled
in different (perhaps opposing) directions: most of the contributions
cancel, giving rise to a small residual force.

A naive approach would compute the contribution at both sides of the
object separately and then add up. This, however, would yield a huge
variance for a comparably small resulting value. Fortunately, the
force contribution of a loop and its mirrored image are highly
correlated in these situations. Often, one is the negative value of
the other. So we have a better way of estimating the contribution by
estimating the expectation of the following process:

\begin{quote}
Randomly pick a loop and also consider its mirror image under the
symmetry; then add up the force contributions of both these loops.
\end{quote}

In that way, we do not change the expectation value of the sum, but,
due to the correlation, the variance is much smaller. In that way, it
is possible to compute force contributions where a naive approach
would require excessive effort due the huge variation, as in the
system discussed in Section~\ref{sec:cylinders}.

\subsection{Parallelization}

In the Worldline formalism the calculation of contributions to the
total energy (or force) can be performed independently for each grid
point. Also, for each grid point, the contribution of each loop to the
energy (or force) does not depend on the contribution of the other
loops. Thus the problem is easily seen as being embarrassingly
parallel, and furthermore the basic component -- processing a loop --
does not require overy complex calculations (in the sense of memory
requirements and algorithmic effort). As the worldline method is a
probabilistic approach, the accuracy of the calculation can be
increased (within reasonable limits dictated by computational effort)
by increasing the number of grid points, number of loops, and the
number of points per loop in an appropriate way.

As the computation for each point and loop follows the same algorithm,
this approach fits the \textsmaller{SIMD} (single instruction,
multiple data) processing approach very well. Powerful massively
parallel \textsmaller{SIMD} hardware is now available at a highly
competitive price in the form of specialized Graphics Processing Units
(\textsmaller{GPU}s), where development has -- to a large extent --
been driven by the video game industry.

When implementing Worldline Numerics on \textsmaller{GPU} hardware,
one has to bear in mind certain constraints of
\textsmaller{GPU} programming. The memory hierarchy of most
\textsmaller{GPU}s discerns between global memory and a special form
of fast local memory called ``shared memory''. It is often favourable to perform most of the
computations in shared memory. This fast shared memory typically is
quite small and has to be divided between simultaneously running work
items, thus limiting the number of loop points that can be computed
concurrently. One also has to keep in mind that \textsmaller{GPU}s
perform rather poorly on complex branching patterns and the hardware
optimizations on \textsmaller{GPU}s mainly target a high throughput of
floating point operations, often neglecting the performance on integer
calculations needed for most Pseudo Random Number Generators
(\textsmaller{PRNG}s).

On account of shared memory size limitations, the authors developed a
version of the d-loop algorithm that generates, for each loop, the
loop points on the fly -- without ever storing the entire loop in
memory. As \textsmaller{GPU}s do not support recursion directly, it is
advantageous to instead manage a stack of arcs yet to be split in
half, as sketched in figure~\ref{fig:gpustack}.

\begin{figure}

\begin{center}\begin{minipage}{0.7\textwidth}{\scriptsize
\begin{lstlisting}
while (stacksize>0) {
      pop(StartPos, EndPos, &level);
      if (level > 0){
            MidPos = (StartPos + EndPos) / 2 + gauss_normal(0, sigma(level));
            push(MidPos, EndPos, level-1);
            push(StartPos, MidPos, level-1);
            }
      else calc_contribution(StartPos);
}
\end{lstlisting}}
\end{minipage}
\end{center}
\caption{\scriptsize Schematic structure of the code managing a stack of 
yet-to-be-split arcs.}
\label{fig:gpustack}
\end{figure}

This manages to reduce the memory footprint of loop generation and
processing from $\mathcal{O}(N)$ to $\mathcal{O}(\log(N))$, $N$~being
the number of loop points. This makes it possible in principle to
shift loop generation to \textsmaller{GPU} cores even for loops too
large to fit into the shared \textsmaller{GPU} memory. In most cases,
however, it seems more appropriate to instead pre-generate loop shapes
on the \textsmaller{CPU} and subsequently upload them
into \textsmaller{GPU} read-only memory visible to each \textsmaller{GPU} work
item. The \textsmaller{GPU} threads then perform intersection checks
for loops of pre-determined shape shifted to different grid points.
Using the same set of loops at all grid points also helps in terms of
statistics as there then can be direct cancellation between opposing
forces arising from similar geometric structures. (See also the
discussion in Section~\ref{subsubsec:cancellation}.)

The authors so far used \textsmaller{GPU}s mainly with the direct numerical
integration method described in section~\ref{subsec:directintegration}
and have at the time of this writing not yet generalized the
(algorithmically much less uniform) handling of scale factor intervals
to the \textsmaller{GPU} for different geometries. First computations
in the plate-plate geometry show a much better convergence than a
full Monte-Carlo integration.

As language for the implementation of the algorithm the authors have
used \textsmaller{OpenCL} in combination with \textsmaller{PyOpenCL}\cite{pyopencl}. 
This helps code re-use across a broad spectrum of multi-~or manycore architectures.

\subsection{Generalization to Electrodynamics?}
While Worldline Numerics has drawbacks in comparison to other
computational approaches, such as e.g. its inability to easily
incorporate frequency-dependent optical properties of immersion media,
it also has, from an applications perspective, some potentially very
attractive characteristics (as we have reasoned out in this article).
The biggest present obstacle to the utilization of Worldline Numerics
for engineering applications is that it has not yet been generalized
from (massless and massive) scalar fields to photons interacting with
conductors. As we demonstrate in section~\ref{sec:cylinders}, trying
to use scalars to approximate the behaviour of photons can easily give
qualitatively wrong results -- so, the need to be able to handle
photons is quite pressing from an application perspective.

While worldline methods can in principle be adopted to dealing with
vector bosons, as has been explored e.g. in~\cite{2001PhR...355...73S}
for gluon loop radiative corrections, the challenge is in properly
modeling the boundary conditions for photon-conductor
interaction. Considering the structure of the theory, one would at
least expect that it should be achievable to couple the photon to a
charged scalar undergoing spontaneous symmetry breaking, giving it an
effective mass inside each object. This would then amount to modeling
electromagnetic Casimir forces in the presence of superconductors.

Rather than trying to construct a generalization of Worldline Numerics
for photons and conductors starting from quantum electrodynamics
principles, we for now approach this problem in a more adventurous
way, if only to generate ideas: here, our guiding question is what the
simplest conceivable generalization of Worldline Numerics might be
that could possibly stand a chance of modeling electrodynamics.
Naturally, the subsequent discussion in this section will be of highly
speculative nature.

Obviously, such a generalization will involve having photons propagate
in loops. As we are not particularly interested in manifest Lorentz
invariance here, we may just as well straightaway choose to gauge away
the non-dynamic timelike component of the electromagnetic vector
potential $A_0$, leaving us with a three-dimensional vector describing
the photon polarization state. This is, of course, still redundant, as
the photon only has two (transversal) physical states rather than
three, but it might be conceivable to ultimately end up with a
formulation in which, after evaluating the integrals over loop shapes
and sizes, it turns out that the boundaries do not interact with
(i.e. never see) the longitudinal photons. In such a case, they would
drop out from all Casimir forces.

It makes sense to assume that, as in the scalar case, bodies can be
modeled as hollow thin shells; if we accept the perfect conductor
approximation (which in particular states that characteristic length
scales from the geometry are large in comparison to characteristic
atomic distances), we would not expect to be able to probe the inner
structure of conductors by looking at Casimir forces between them. So,
we need to only concern ourselves with what happens when a photon loop
pierces a conductor surface. The boundary conditions of a perfect
conductor ($\vec n\times\vec E=0$ -- no electric field parallel to
surface, as this would immediately be compensated by charges shifting
accordingly, and $\vec n \cdot\vec B=0$ -- no magnetic field
perpendicular to surface, as this would be compensated by eddy
currents, at least for frequencies $\omega$ sufficiently large for the
perfect conductor approximation to hold) have to be implemented in
some way. We want to consider all possible (including un-physical
longitudinal) photon polarizations simultaneously, and hence should
associate to every loop edge a $3\times 3$ matrix acting on the
polarization state. In the end, the contribution to the Casimir energy
from the loop under consideration should be taken to be the trace of
the product of all these matrices; free propagation will amount to the
identity, and the matrix corresponding to an edge that pierces a
conductor surface would involve a projection eliminating some
polarizations. What form may such a projection matrix have? As we
already accounted for photon polarization directions, the only
directions available are the surface normal~$\vec n$, as well as the
local velocity~$\dot{\vec{ y}}$. The 
$\vec E=\partial_t A\sim\omega \vec A$ condition only involves
perpendicularity to $\vec n$, and while the $\vec B$-condition would
be expected to pick up spatial components of $\dot{\vec y}$,
re-scaling these in order to form a projector seems to also bring us 
to a $\vec A\perp \vec n$ condition, which means that one should take
as projector that restricts to forbidden states the 
matrix~$I-\vec n\otimes\vec n$. Evidently, in the case of parallel 
plates, the trace would just introduce a factor two relative to 
the scalar field, as desired.

\begin{figure}
\centerline{\includegraphics[width=0.8\textwidth]{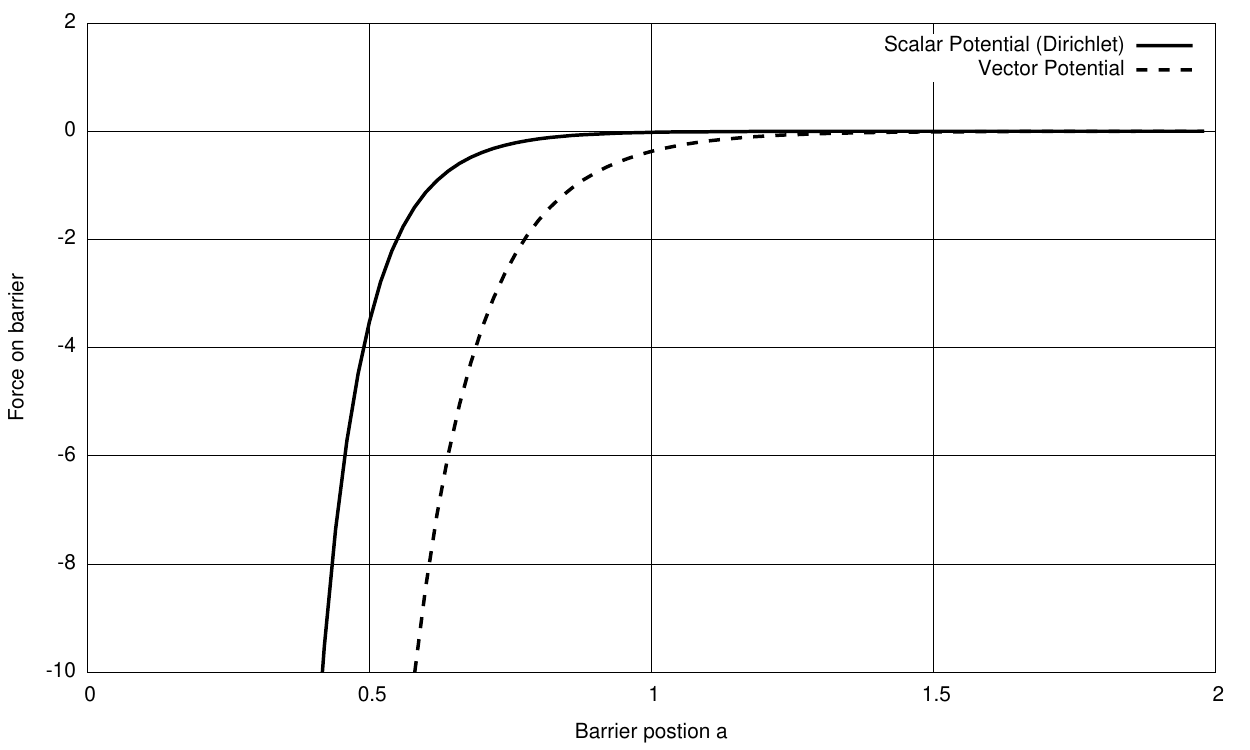}}
\caption{\scriptsize Casimir forces in the piston geometry as a function of barrier height, 
calculated for scalars and photons using the analytic expressions
given in~\cite{2007PhRvD..76d5016H}.}
\label{fig:exact}
\end{figure}

\begin{figure}
\begin{center}
\centerline{\includegraphics[width=0.8\textwidth]{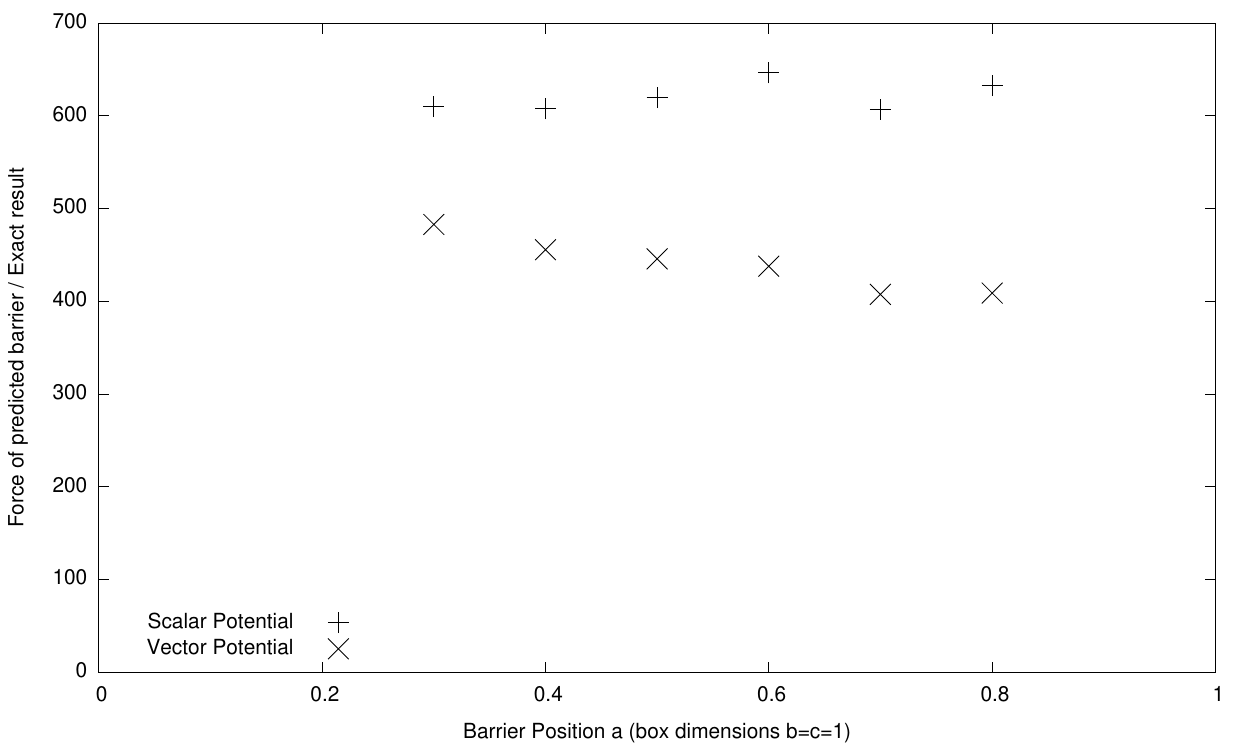}}
\end{center}
\caption{\scriptsize Comparison of predicted forces in the Casimir Piston geometry: 
The graph shows the ratio of the Casimir forces computed 
with our method to the exact, analytically computed,
values~\cite{2007PhRvD..76d5016H}, see
Figure~\ref{fig:exact}. This ratio is plotted for
computations based on scalar fields (upper curve) and photons (lower
curve). A constant ratio would indicate perfect coincidence of the
methods; note that constant scaling factors (e.g., due to fundamental
constants of physics) have not been taken into account in our
computations, as they were only carried out to test the functional
dependency.}
\label{fig:EM-Casimir}
\end{figure}

The prescription described above seems, at the superficial level,
quite obviously wrong -- this is most easily seen by considering a
loop close to a box corner made up of different objects: if it pierces
three mutually perpendicular faces, it would contribute three
forbidden polarizations, where the physical photon only has
two. Still, it is valid to ask whether this loop processing
prescription is related to some idealization of boundaries in a
quantum field theory of vectors -- and ask what it might look like.

One would, for the reason given above, naturally expect such a
prescription to give energies that do not have much in common with
photon Casimir energies. An attractive nontrivial test geometry for
which scalar and photon Casimir forces are known exactly, where
discrepancies could reasonably be expected to manifest, and in which
this scheme is easily implemented, is the piston geometry studied
in~\cite{2007PhRvD..76d5016H}. If we plot the quotient of the (not yet
properly normalized) energies predicted by Worldline Numerics and the
analytic result for the scalar field, we get curve~1 in
figure~\ref{fig:EM-Casimir}. If we compute the same ratio for our
aventurous ``Vector Worldline'' generalization and the known result
for photons, we get curve~2.

We consider a coordinate aligned closed box with square cross section
of side length~$b=c=1$ and height height~$h\gg b$ -- here, we
choose~$h=100$. This box is sub-divided by a coordinate-aligned
box-shaped movable barrier of identical cross section and infinitely
small thickness; the distance between barrier and base of the box is
taken to lie in the interval~$[0;1]$.

The spatial sampling of loop positions was done on a Cartesian grid
with lattice spacing~$0.05$, covering a coordinate-parallel cube of
edge length~$3$ centered at the point 0.5 above the center of the box
base plate. Forces have been calculated for barrier heights from~$0.3$
to $0.8$ in steps of~$0.1$, using a central difference quotient with
$\delta h=0.1$.

Since the barrier is taken as being as wide as the box, there is no
positive minimum to the contributing loop size. However, for loops
much smaller than the distance between barrier and base of the box the
local geometry on either side of the barrier symmetric. So the
contribution of sufficiently small loops cancels out. In order to
avoid infinities, the loop scaling factor~$s$ was constrained to the
range~$[0.02;6]$ and sampled as described in
Section~\ref{subsec:directintegration}.

In this system, one must keep in mind that, for the distances
involved, Casimir energies, both in the scalar and photon case, vary
by more than three orders of magnitude -- as shown in
Figure~\ref{fig:exact}. While the ratio clearly shows some drift for
the photon case in this experimental calculation, it cannot yet be
excluded that this may be a discretization related artefact. Given
that the simplistic ansatz proposed here would be expected to quite
clearly show up as being deeply wrong (considering the ratios
involved), this outcome is somewhat remarkable. At the very least, it
seems to make sense to try and explore a small number of other
geometries to learn whether this is a curious coincidence for this
particular geometry (as one would expect), whether some not yet
understood crazy cancellations may actually make this method work as a
model for photons (unlikely), or whether it may turn out as being
somewhat useful pragmatically in the sense of a heuristic engineering
method that is known to be mathematically flawed (as many such
engineering heuristics are), but manages to give a reasonably good
estimate for some applications.

\section{Parallel Cylinders between Plates, revisited}\label{sec:cylinders}

\begin{figure}
\centerline{\includegraphics[width=0.3\textwidth]{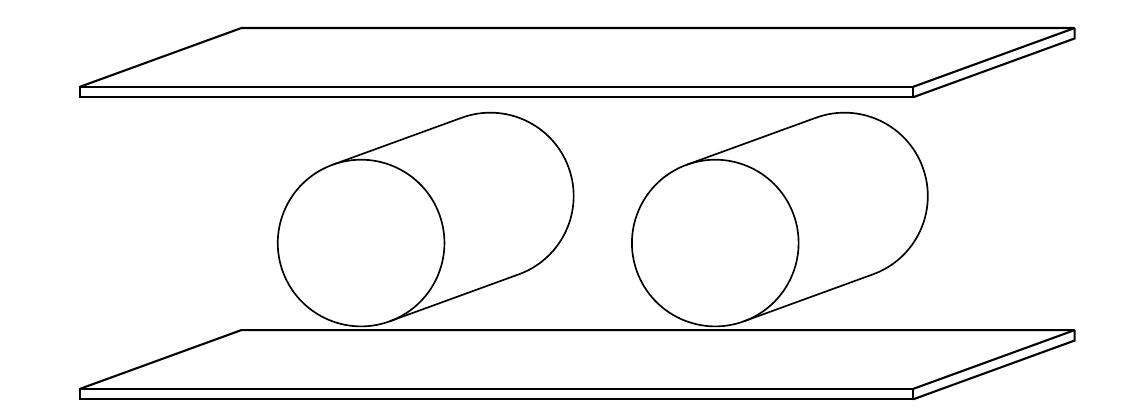}%
            \includegraphics[width=0.6\textwidth]{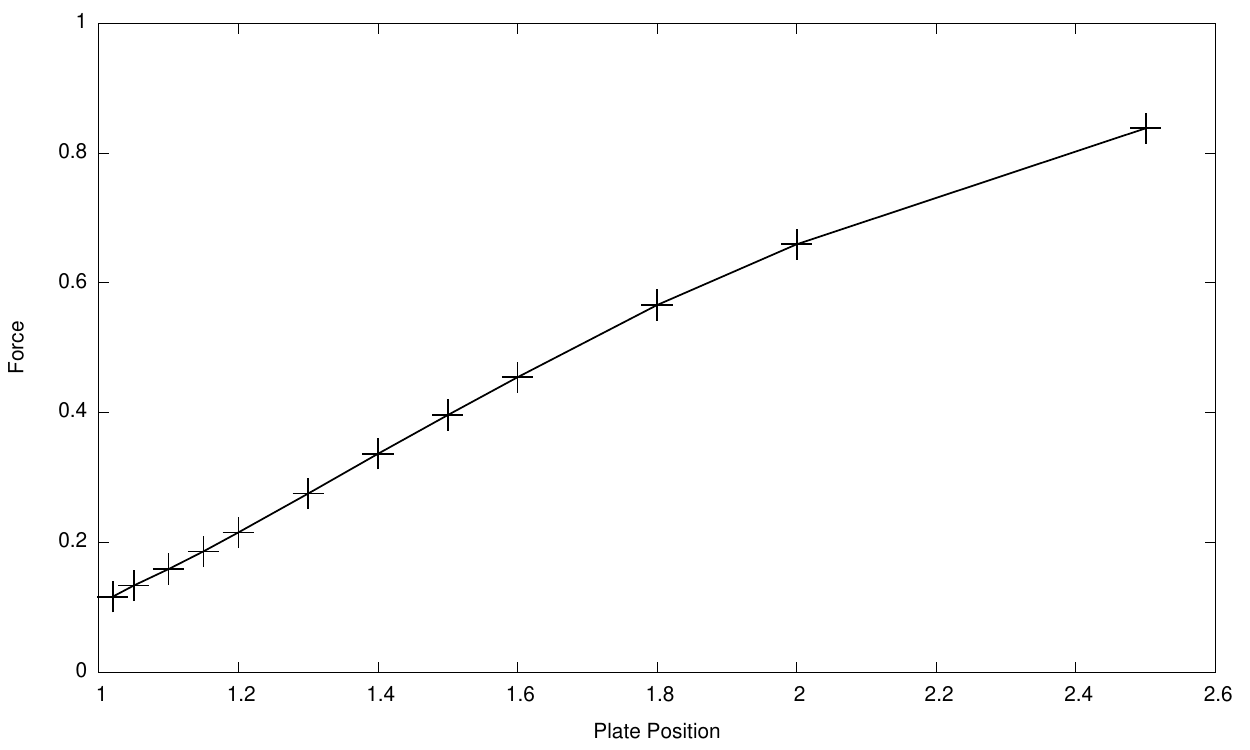}}
\caption{\scriptsize The ``cylinders with sidewalls'' geometry and the dependency
of the attractive scalar Casimir force between the cylinders on the Position of the
plates. At position 1.0, the plates would touch the cylinders.}
\label{fig:cylindersplates}
\end{figure}

The ``cylinders with sidewalls'' geometry studied
in~\cite{2008PhRvA..77c0101R} (Figure~\ref{fig:cylindersplates}, on
the left) has been shown to nicely demonstrate that Casimir forces are
essentially multi-body forces. It is given by two parallel, infinitely
long cylinders between two parallel infinite plates. Focusing on the
attractive force between cylinders, one finds that this depends in a
fairly subtle way on the distance between the plates.

For the calculation whose results are shown in
figure~\ref{fig:cylindersplates}, the cylinders used have radius 1.0
and centres at ($x$, $y$)-coordinates (-2.0, 0.0) and (2.0, 0.0),
respectively.  The plates are given by the equation $z=p$, where $p$
is varied in the range 1.02\ldots{}2.50. The calculation of the force
on the left cylinder used the scaling method, sampling around a grid
with spacing 0.05 and exploiting the mirror method
(Section~\ref{subsubsec:cancellation}) by taking a reflection-symmetric
loop ensemble w.r.t. the plane $x$=-2.0 to reduce the variance due to
cancellation. This sampling was done in an adaptive manner
(Section~\ref{subsec:adaptive}). In total, between $1.9\cdot 10^7$ and
$3.9\cdot 10^7$ loops of $2^{13}$ points were sampled for each geometry.

The results are shown at the right-hand side of
Figure~\ref{fig:cylindersplates}.  As opposed to methods such as the
proximity force approximation, we do see a dependency of the forces on
the cylinders on the plate distance. We however could not find the
non-monotonic behaviour reported in the
literature~\cite{2008PhRvA..77c0101R}. So, once again, we have an
example where scalars behave in a qualitatively different way than
photons~\cite{2007PhRvD..76d5016H}.

\section{Conclusions}

From a microsystems engineering perspective, the Worldline Numerics /
Loop Cloud Method has a number of attractive properties, such as the
ability to quickly give crude estimates, considerable potential to
solve geometric optimization related problems, and of course its
conceptual simplicity and intuitiveness that make it a useful
educational tool with the potential to give a simple yet
quantitatively correct mental model of the origin of Casimir forces.
Quite remarkably, the operational procedure can be explained using
very simple concepts only -- in fact, even without having to use much
linear algebra.

At present, the biggest obstacle to its utilization for engineering
applications is the method's inability to handle photons in the
presence of conducting boundaries instead of scalar particles in the
presence of Dirichlet boundary conditions. As we have demonstrated in
section~\ref{sec:cylinders} through an example calculation, the
problem is that any attempt to use scalars in order to approximate
photon Casimir forces is questionable as this can easily give
predictions that are wrong already at the qualitative level.

In addition to this calculation, and to discussing some methodological
improvements of the Loop Cloud Method, this article provided some
early speculation on whether a simple extension of the method may be
able to give useful estimates for photon Casimir forces; despite
obvious conceptual issues, the accompanying computation turned out to
produce numbers that match the known exact result remarkably well, in
particular considering that Casimir forces range over a few orders of
magnitude for the problem studied. Still, one naturally would next try
to refute the viability of this simplistic approach by checking its
predictions for another not too trivial geometry, before starting any
attempt to formally prove its validity.

\paragraph{Acknowledgments}

It is a pleasure to thank Holger Gies and Roman Zwicky for useful
discussions and comments on this subject. \textsmaller{CPU} calculations have been
performed on the University of Southampton's Iridis3 cluster and the
\textsmaller{LOEWE-CSC} cluster of the Goethe Universit\"at
Frankfurt. \textsmaller{GPU} calculations have been performed on the
\textsmaller{LOEWE-CSC} cluster of the Goethe Universit\"at
Frankfurt. J. Gerhard received funding for this project by the
HGS-HIRe Abroad programm of the \textit{Helmholtz Graduate School for Hadron
and Ion Research}.
Financial support by the \textsmaller{UK} Engineering and Physical
Sciences Research Council under \textsmaller{EPSRC} Grant
\textsmaller{EP/H049924/1} is gratefully
acknowledged.

\bibliographystyle{hunsrt}

\end{document}